\documentclass[12pt]{article} 
\usepackage{bbold}
\usepackage{cite}
\usepackage{graphicx}
\def \bea{\begin{eqnarray}} 
\def \beq{\begin{equation}}

\def \eea{\end{eqnarray}} 
\def \eeq{\end{equation}}

\def\lsim{\mathrel{\rlap{\lower3pt\hbox{$\sim$}}\raise2pt\hbox{$<$}}}
\def\gsim{\mathrel{\rlap{\lower3pt\hbox{$\sim$}}\raise2pt\hbox{$>$}}}

\def\bra#1{\left\langle #1\right|}
\def\ket#1{\left| #1\right\rangle}

\begin{document} 
\begin{flushright}
EFI 15-22 \\
TECHNION-PH-2015-09 \\
arXiv:1507.03565 \\
October 2016 \\
\end{flushright} 
\centerline{\bf TEST FOR EXOTIC ISOSCALAR RESONANCE}
\centerline{\bf  DOMINATING $D^0 \to \pi^+\pi^-\pi^0$ DECAYS}
\medskip
\centerline{Michael Gronau}
\centerline{\it Physics Department, Technion, Haifa 32000, Israel}
\medskip 
\centerline{Jonathan L. Rosner} 
\centerline{\it Enrico Fermi Institute and Department of Physics,
  University of Chicago} 
\centerline{\it Chicago, IL 60637, U.S.A.} 
\bigskip

\begin{quote}                                 
The decay $D^0 \to \pi^+ \pi^- \pi^0$ appears to be dominated by $\rho \pi$
states in a configuration of zero total isotopic spin.  The spin $J$, parity
$P$, and charge-conjugation eigenvalue $C$ of this final state are therefore
$J^{PC} = 0^{--}$, which cannot be formed of a quark $q$ and antiquark $\bar q$.
If a resonance near $M(D^0)$ dominates the final state, it must be a {\it
hybrid} composed of a quark-antiquark pair and a constituent gluon, or a {\it
tetraquark} $q q \bar q \bar q$. A test for this resonance in electroproduction
is proposed.
\end{quote}

\leftline{\qquad PACS codes: 12.39.Mk, 13.25.Ft, 13.60.Le, 14.40.Rt}
\bigskip

\section{INTRODUCTION} \label{sec:intro}

The decay of the charmed meson $D^0$ into $\pi^+\pi^-\pi^0$ exhibits a curious
dominance by the state of zero total isotopic spin \cite{Gaspero:2008rs,%
Gaspero:2010pz}.  Since this three-pion state has odd $G$-parity and $I=0$,
its charge-conjugation eigenvalue $C$ is negative.  Since it is a three-pion
state in a state of zero total angular momentum $J$, its parity $P$ is also
negative.  It thus has $J^{PC} = 0^{--}$, a CP-even configuration which cannot
be formed of a quark and antiquark.  The even CP property has been confirmed by
subsequent analyses \cite{Nayak:2014tea,Malde:2015mha}.  The latest finds the
three-pion state to have $CP=+$ $(97.3 \pm 1.7)\%$ of the time, which includes
a small (few-\%) contribution from $I=2$ \cite{Bhattacharya:2010id}. 
This observation has a useful implication for
a precise determination of the CP-
violating CKM (Cabibbo-Kobayashi-Maskawa) phase $\gamma$ in decays of the class
$B^{\pm,0} \to D_{\rm CP}K^{(*)\pm,0}$~\cite{Gronau:1990ra,BDK}. 

The dominance of $I=0$ can be reproduced \cite{Bhattacharya:2010id} in
flavor-SU(3) analyses of all $PV$ decays of charmed mesons, where $P$ and
$V$ stand for light pseudoscalar and vector mesons.  Topological amplitudes
$T$ (``color-favored tree''), $C$ (``color-suppressed tree''), and $E$
(``exchange'') cooperate in such a way as to give $I=0$ intensity fractions
of $(92.9 \pm 6.7)\%$ in the fit of Ref.\ \cite{Bhattacharya:2008ke} and
$(90.9 \pm 18.2)\%$ in the fit of Ref.\ \cite{Cheng:2010ry}.  The possibility
of dominance by a non-$q \bar q$ resonance near $M(D^0) \simeq 1865$ MeV
was mentioned in Refs.\ \cite{Gaspero:2010pz,Bhattacharya:2010id}.  In the 
present work we propose a means of testing this hypothesis.

This paper is organized as follows.  In Section \ref{sec:pr} we review some
properties of resonances in charm decays and of a new hybrid state near $M_D$.  
We then stress the need for electroproduction of a spinless resonance via pion
exchange in Sec.\ \ref{sec:ele}.  Possible interpretations of a signal are
described in Sec.\ \ref{sec:int}.  A detailed program for anticipating signal
strength is set forth in Sec.\ \ref{sec:sig}, while possible variants to this
approach are noted in Sec.\ \ref{sec:var}.  Sec.\ \ref{sec:sum} summarizes.
An Appendix examines assumed relations among photoproduction amplitudes
of light mesons.

\section{RESONANCES IN CHARM DECAYS} \label{sec:pr}

The role of nearby resonance states in charmed meson decays has been pointed 
out a long time ago~\cite{Lipkin:1980es}.  Dominant contributions to $D^0 \to
K^-\pi^+$ of strangeness $-1$ $q \bar q$ 
resonances with masses below and near the $D^0$ mass have been studied in 
Ref.~\cite{Gronau:1999zt}. It was also argued in Ref.~\cite{Cheng:2010ry} that 
an intermediate glueball state at $f_0(1710)$ could explain the large ratio 
$\Gamma(D^0 \to K^+K^-)/\Gamma(D^0\to \pi^+\pi^-)$.

We denote the proposed resonance by $X(0^{--})$, where the quantity in
parentheses refers to $J^{PC}$.  The Dalitz plot for $D^0 \to \pi^+ \pi^-
\pi^0$ appears to be dominated by three $\rho \pi$ bands of approximately
equal strength; they would be strictly equal in the $I=0$ limit.

A contribution of a given resonance $R$ to $D^0$ decay into a final state $f$
is given by
\beq
A_R(D^0 \to f) = \frac{\langle R| H_W| D^0\rangle g_{Rf}}
{m^2_D - m^2_R - im_R\Gamma_R}~,
\eeq
where $\langle R |H_W| D^0\rangle$ is the weak Hamiltonian matrix element
between $D^0$ and $R$ states, $g_{Rf}$ is the strong decay coupling of the
resonance to $f$, while $m_R$ and $\Gamma_R$ are the resonance mass and width.
We will now compare the two factors in the numerator and the Breit-Wigner
denominator for $R=X(1865), f=\rho\pi$ and $R=\bar K^{*0}(1430), f = K^-\pi^+$.

It seems impossible to present a reliable model for calculating the strong
coupling (or the width) of the hybrid meson $X(0^{--})$ to $\rho\pi$.  No such
attempt has been made in Refs.\ \cite{Ketzer:2012vn}--\cite{QSSR} studying
hybrid states in QCD. This stands in contrast to the strong coupling of the
strangeness $-1$ $q \bar q$ spin zero resonance $\bar K^{*0}(1430)$ to
$K^-\pi^+$ which has been well measured through the resonance width \cite{PDG}.
The contribution of $\bar K^{*0}(1430)$, peaking 436 MeV below the $D^0$ mass,
to the $D^0 \to K^-\pi^+$ amplitude has been calculated to be around $30\%$ of
this amplitude \cite{Gronau:1999zt}. The latter paper also suggested that
another $s \bar d$ $P$-wave resonance (most likely an $n=2$ radial excitation)
around 1900 MeV may dominate the amplitude.

It is difficult to compare quantitatively $D^0$ weak interaction matrix
elements for final states $|X(1865)\rangle$ and $|\bar K^{*0}(1430)\rangle$
involving CKM factors $V_{cd}V^*_{ud}$ and $V_{cs}V^*_{ud}$. The respective
quark transitions $c (\bar u)\to d \bar u d (\bar u)$ and $c (\bar u) \to s
\bar u d (\bar u)$ involve two quark-antiquark pairs in the final state. This
seems to favor a tetraquark state $X$ in the first transition over a
quark-antiquark state $\bar K^{*0}$ in the second transition.

Considering only the magnitudes of the two Breit-Wigner denominators for $D^0
\to X(1865) \to \rho \pi$ and $D^0 \to \bar K^{*0}(1430) \to K^-\pi^+$ one
finds their
ratio to be $2.08/(1.865\,\Gamma_X)$~\cite{PDG}, where $\Gamma_X$ is given in
GeV.  For $\Gamma_X=0.3$ GeV this by itself would favor by a factor 3.7 this
resonant contribution to $D^0 \to \pi^+\pi^-\pi^0$ over the contribution of
$\bar K^{*0}(1430)$ to $D^0 \to K^-\pi^+$, and thus favor essentially complete
$X(1865)$ dominance of the former decay.

\section{ELECTROPRODUCTION IN PION EXCHANGE} \label{sec:ele}

\begin{figure}
\begin{center}
\includegraphics[width=0.8\textwidth]{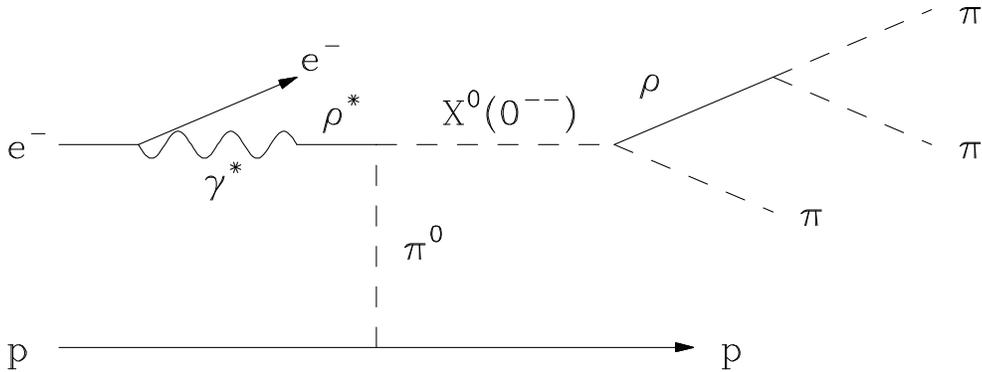}
\end{center}
\caption{Electroproduction of a hypothetical $X(0^{--})$ resonance with mass
near $M(D^0 \simeq 1865$ MeV, observed through its decay to $\pi^+\pi^-\pi^0$.
\label{fig:phot}}
\end{figure}

Based on the coupling of the resonance to $\rho^0 \pi^0$, we propose to
electroproduce it on a proton target via $\pi^0$ exchange, as illustrated in
Fig.\ \ref{fig:phot}.  Photoproduction of a spinless state off a nearly real
pion target is a forbidden $0 \to 0$ electromagnetic transition.  Hence the
cross section should vanish as the squared momentum transfer $q^2$ goes to
zero.  This behavior is familiar from the two-photon reaction $e^+e^- \to e^+
e^- f_1(1285)$ \cite{Aihara:1988uh,Gidal:1987,Achard:2001uu}. Here the
excitation of the spin-1 $f_1$ by two real photons is forbidden by the
Landau-Yang theorem \cite{Landau,Yang}, so at least one of the electrons must
undergo significant recoil.  Pion exchange is most effective at small momentum
transfers \cite{Chew:1958wd,Williams:1970rg}, so virtual photons of the highest
possible energy have an advantage in producing a massive state.

There will be conventional $q \bar q$ resonances coupling to $\rho \pi$.  At
lower masses these include $a_1(1260)~(J=1)$, $a_2(1320)~(J=2)$, $\omega(1650)
~(J=1)$, and
$\omega(1670)~(J=3)$ \cite{PDG}.  However, the $X(0^{--})$ should have a
distinctive signature.  It should decay mainly to $\rho \pi$, populating
each of the three $\rho \pi$ bands equally, with a characteristic null along
all three symmetry axes of the Dalitz plot \cite{Zemach:1964}.  Furthermore,
as mentioned, its production via pion exchange should be suppressed as the
virtual photon becomes closer to the mass shell.

\begin{figure}
\begin{center}
\includegraphics[width=0.9\textwidth]{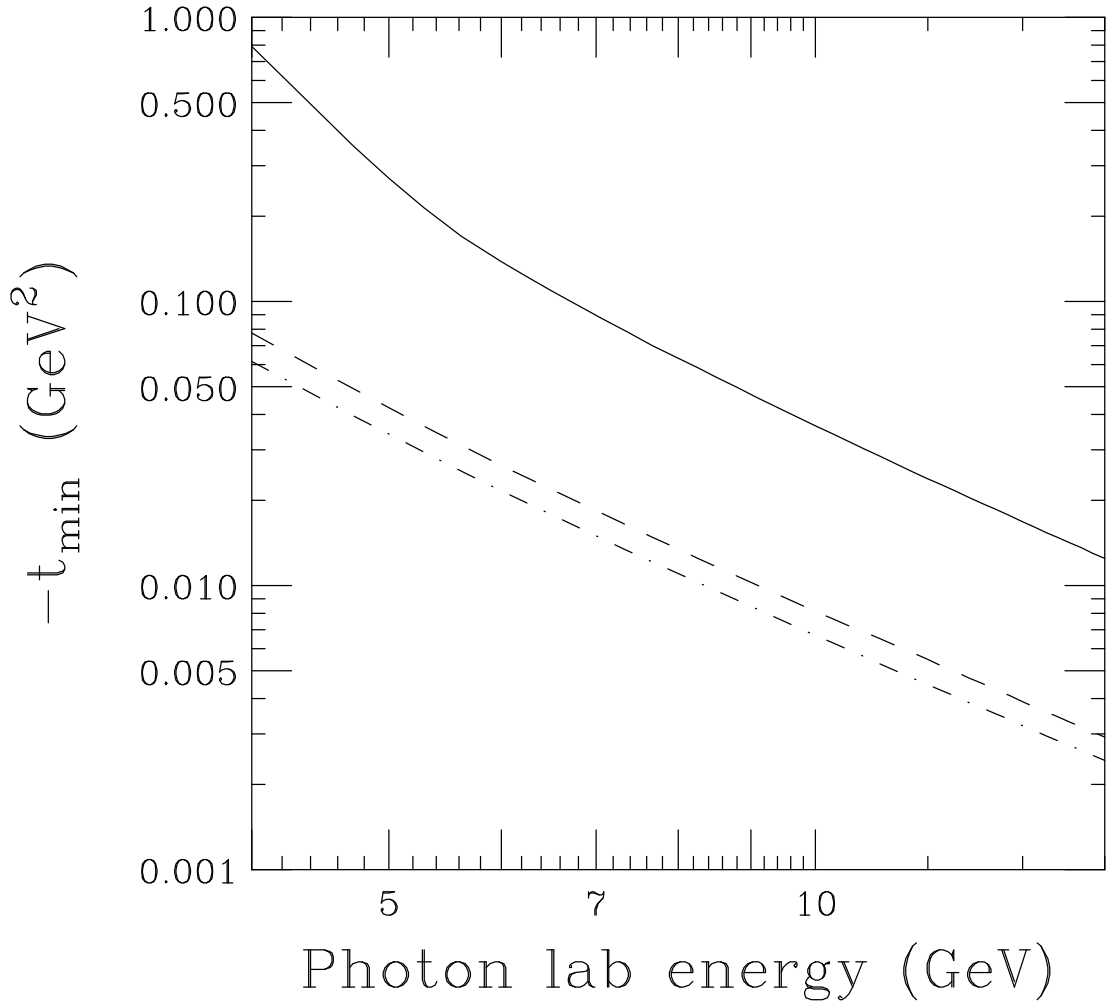}
\end{center}
\caption{Values of $-t_{\rm min}$ for $\gamma p \to X p$ as functions of
incident photon laboratory energy $E_\gamma$.  Solid line: $X = X(0^{--})$
[$M(X) = M(D^0 = 1865$ MeV]; dashed line: $X = a_2(1320)$; dot-dashed line:
$X = a_1(1260)$.  For large $E_\gamma$, $-t_{\rm min} \simeq M_X^4/(4
E_\gamma^2)$.  When the photon is virtual with $q^2 < 0$ this expression
becomes $[(M_X^2 - q^2)/(2E_\gamma)]^2$.
\label{fig:tmin}}
\end{figure}

The minimum momentum transfer should be of order $-m_\pi^2$ or smaller to
efficiently utilize the pion pole.  As shown in Fig.\ \ref{fig:tmin}~\cite{PDG},
photons of 6 GeV (the original energy at Jefferson National Laboratory [JLAB])
can achieve $|t_{\rm min}| = {\cal O}(m_\pi^2)$ when exciting $a_1(1260)$ or
$a_2(1320)$, while at least $E_\gamma = 12$ GeV (the upgraded JLAB energy) is
required to achieve sufficiently small $|t_{\rm min}|$ when exciting a state
with mass $M(D^0)$.  Tagged photons in the 4.8--5.4 GeV range have been used
by the CLAS Collaboration at JLAB to photoproduce $a_2$ and $\pi(1670)$
\cite{Nozar:2008aa}, but no signal for $a_1$ was seen.  Pion exchange seems
to account satisfactorily for $a_2$ production in this experiment and others
in the 4--7 GeV range.  (See Fig.\ 3 of \cite{Wang:2015kia}.)  It is noted in
Ref.\ \cite{Wang:2015kia} that the COMPASS experiment at CERN
\cite{Abbon:2007pq}, using muons of energy 160--200 GeV, also is capable of
photoproducing or electroproducing light-quark meson states.

Other production mechanisms besides electroproduction are possible.  For
example, the process $\pi^+ p \to X(0^{--}) \Delta^{++}$ can proceed through
charged $\rho$ exchange, leading to a final state $(\pi^+ \pi^- \pi^0)
(\pi^+ p)$. Photoproduction of an $X(0^{--})$ can receive nonzero contributions
from exchange of any neutral meson with $J \ne 0$ and $C = +$, such as the
$a_1$ or $a_2$.

\section{INTERPRETATION OF A SIGNAL} \label{sec:int}

If the resonance is seen, it could be a hybrid or a tetraquark.  A $0^{--}$
hybrid occurs in models involving constituent gluons \cite{Ketzer:2012vn,cg1,%
cg2}.  In Ref.\ \cite{cg1} it is expected in the mass range 1.8--2.2 GeV, so it
could dominate $D^0$ decays.  It is predicted to lie somewhat higher (2.3 GeV)
in Ref.\ \cite{cg2}.  A gluon
with $J^{PC} = 1^{--}$ can combine with a color-octet $I=0$ $q \bar q$
state with $J^{PC} = 1^{++}$ (i.e., a $^3P_1$ state) in a state of relative
orbital angular momentum zero to form the $0^{--}$ hybrid.  Lower-mass
hybrids in the range 1.3--2.1 GeV can be formed with a gluon and a color-octet
$^1S_0$ or $^3S_1$ $q \bar q$ state, leading to hybrids with $J^{PC} = 1^{+-}$
and $(0,1,2)^{++}$, respectively.  None of these is exotic.  The lowest-lying
exotic, with $J^{PC} = 1^{-+}$, is predicted in the constituent-gluon model
to lie in the mass range 1.8--2.2 GeV, and to be composed of a gluon and
a $q \bar q$ state with $J^{PC} = 1^{+-}$.  Candidates for this state have
shown up, typically at lower mass \cite{Ketzer:2012vn}.  It is interesting that
other models considered in Ref.\ \cite{Ketzer:2012vn} (bag \cite{bag}, flux
tube \cite{ft}, lattice QCD \cite{LQCD}, QCD spectral sum rules \cite{QSSR})
do {\it not} predict a $0^{--}$ state at comparable mass.

A tetraquark could serve as a proxy for a model with a constituent gluon,
by the simple substitution of a color-octet, spin-1 $q \bar q$ $^3S_1$ pair in
place of the gluon.  Here there are more opportunities for forming a $0^{--}$
state with zero isospin, as both the $^3P_1$ and $^3S_1$ states can have
either zero or unit isospin.  (We are assuming only the two lightest quark
flavors.)

\section{ANTICIPATING SIGNAL STRENGTH} \label{sec:sig}

The forthcoming GlueX experiment \cite{Shepherd:2014hsa} is dedicated to
searching for exotic states of matter in photon-proton collisions.  However,
the detector cannot be operated in an electroproduction mode, so production
of the $X(0^{--})$ should be suppressed.  GlueX {\it should} be able to
photoproduce both $\omega(782)$ and its presumed radial excitation
$\omega(1650)$, important steps (as we shall see below) toward
electroproduction of the $0^{--}$ state.  On the other hand, the CLAS12
detector \cite{Stepanyan:2010kx} is designed to study resonance production with
virtual as well as quasi-real photons, so it should be able to see the
$X(0^{--})$, with cross section decreasing as $q^2 \to 0$.  In the following we
suggest experimental steps to see whether the required sensitivity can be
achieved.

\begin{figure}
\begin{center}
\includegraphics[width=0.92\textwidth]{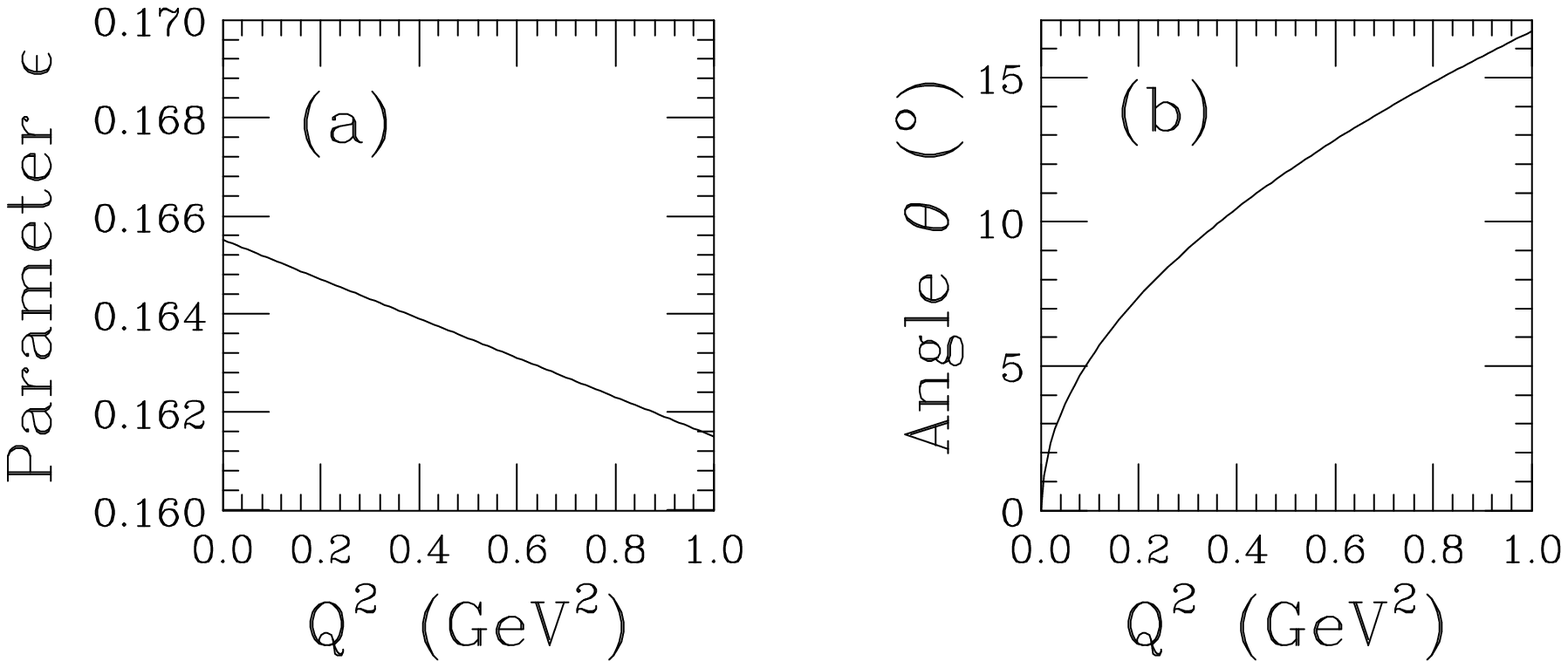}
\includegraphics[width=0.92\textwidth]{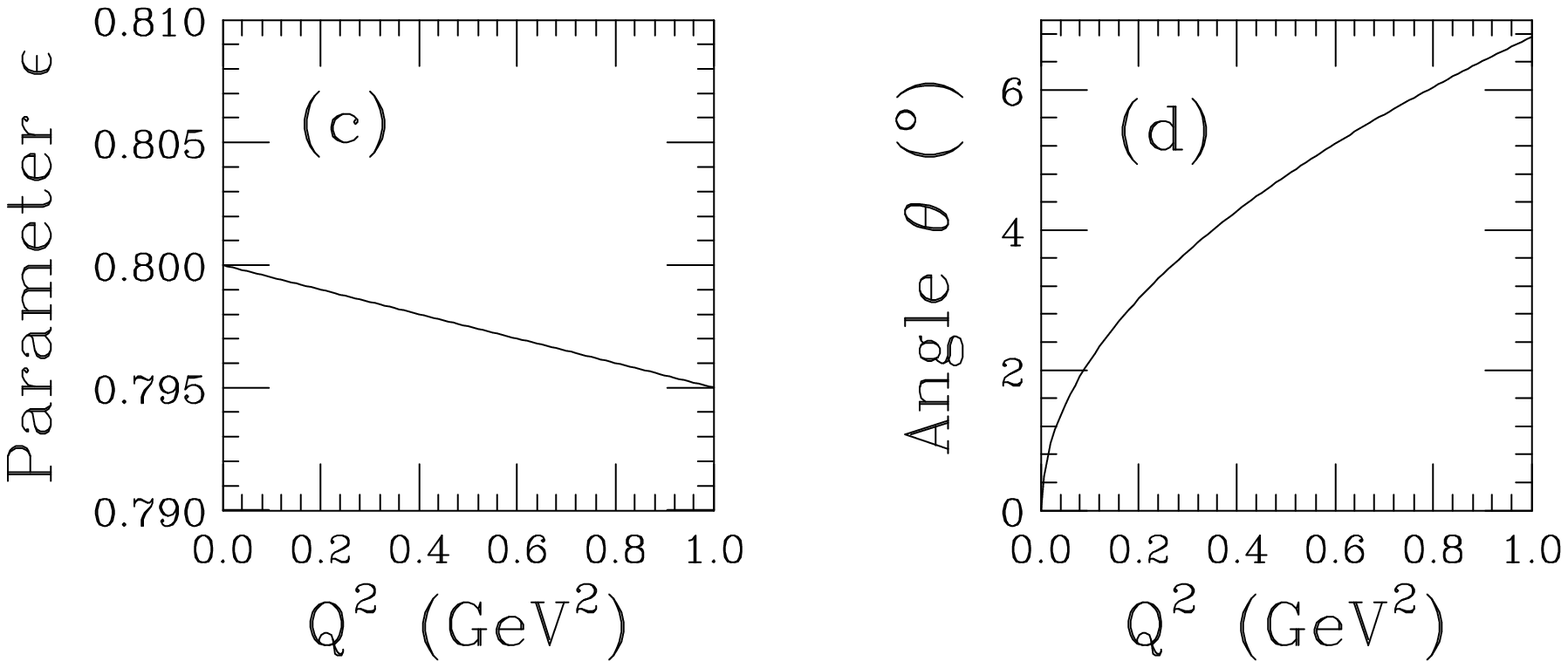}
\end{center}
\caption{Behavior of virtual photon polarization parameter $\epsilon$ (a,c) and
laboratory scattering angle $\theta$ (b,d) as functions of $Q^2$ for $E=12$ GeV
and $E'=1$ GeV (a,b) or 6 GeV (c,d).
\label{fig:eps}}
\end{figure}

The electroproduction of a state whose production by real photons is forbidden
requires one to know the relative flux of longitudinally and transversely
polarized photons $\gamma^*$ produced by a scattered electron:
\beq
e^-(k) \to \gamma^*(q) + e^-(k')~,
\eeq
where $E$ and $E'$ are the laboratory energies of the initial and final
electron, $\nu \equiv E - E'$, and $Q^2 \equiv - q^2 = 4EE'\sin^2(\theta/2)$,
where $\theta$ is the electron scattering angle in the laboratory.
The cross section for $e^- + p \to e^- + $ (anything) can be decomposed into
contributions from transversely and longitudinally polarized virtual photons
\cite{Mamyan:2012th}:
\beq \label{eqn:eps}
\frac{d^2\sigma}{d\Omega dE'} \propto \sigma_T + \epsilon \sigma_L~,~~
\epsilon \equiv \left[1 + 2 \left(\frac{\nu^2}{Q^2} +1 \right)
\tan^2(\theta/2) \right]^{-1}~.
\eeq
One finds the following exact dependence of $\epsilon$ on $E, E'$ and $Q^2$:
\beq
\epsilon = \frac{4EE' - Q^2}{2(E^2 + E'^2) + Q^2}~.
\eeq
Two examples of the behavior of $\epsilon$ and
$\theta$ as functions of $Q^2$ are shown in Fig. \ref{fig:eps} for $E = 12$
GeV and $E'=1$ [3(a) and 3(b)] or $E'=6$ [3(c) and 3(d)].  
For low values of $Q^2$ $\epsilon$ is very weakly dependent on this variable 
and is almost entirely a function of the ratio $E'/E$, 
$\epsilon \simeq 2(E'/E)/[1 + (E'/E)^2]$.

An amplitude for a process forbidden for real photons ($q^2=0$) will behave for
$Q^2 \to 0$ as $Q^2/M_0^2$, where $M_0$ is some characteristic hadron mass.
Taking it to be $m_\rho$, we may expect a suppression by about a factor of 2
relative to a typical photoproduction amplitude when $Q^2 \sim 0.3$ GeV$^2$.
This is the maximum envisioned in one proposed CLAS12 experiment at JLAB
\cite{MBPC}.

If one wants the virtual photon to transfer as much energy $\nu$ as possible
to the hadronic system, one wants $E'$ to be not too large, as in Figs.\ 3(a,b).
However, one pays the price of a smaller factor $\epsilon \simeq 2x/(1 + x^2)$,
where $x = E'/E$, which is about 1/6 for $x = 1/12$.  Roughly speaking, then,
electroproduction of a state that cannot be photoproduced with real photons
will cost about an order of magnitude in cross section relative to a
state that {\em can} be photoproduced.

Now we seek a reference cross section for electroproduction of a {\it known}
state with mass not too different from that of $X(1865,0^{--})$.  We first
look for evidence of $\pi^0$ exchange in {\it photoproduction}.  This will
give rise to neutral states with odd $C$.  Such a process has been seen
in photoproduction of the $\omega(782,1^{--})$ meson with polarized photons
of energy 2.8, 4.7, and 9.3 GeV \cite{Ballam:1972eq}.  The photon polarization
enables the isolation of unnatural parity (i.e., pion) exchange.  Fitting the
differential cross sections $d \sigma/dt$ at all three energies, where $t$ is
the invariant momentum transfer, one finds for unnatural-parity exchange
\beq
\frac{d \sigma^U}{dt}=A e^{bt}/E_\gamma^2~,~~A=164~{\rm nb}~,~~b=7{\rm~GeV}
^{-2}~,
\eeq
yielding $\sigma^U = 0.23$ nb at $E_\gamma = 10$ GeV.  The first stage in
observing electroproduction of $X(1865,0^{--})$ would be to see evidence for
$\pi^0$ exchange in $\omega(782)$ photoproduction.

The next step is to observe $\pi^0$ exchange in photoproduction of a $C=-$
state as close as possible in mass to the $X(1865)$.  Such a state is a radial
excitation $\omega'$ of the $\omega(782)$, denoted in \cite{PDG} by
$\omega(1650)$.  As its mass is quoted as $1670 \pm 30$ MeV, we shall refer
to it as $\omega'$ to avoid confusion with the $J=3$ state of similar mass.
If produced with a virtual photon of small
$Q^2$ and $\nu = 10$ GeV, and with the same differential cross section as
for $\omega(782)$ production, the effect of $t_{\rm min}$ is only suppression
by a factor of 0.85.  However, one can estimate (see the Appendix) using
vector dominance and the known total width of $\omega'$ that a
further suppression factor of $\sim 0.5$ is likely, leading to an overall
suppression factor of about 0.4 and a net cross section of 0.1 nb.

One must then observe the pion-exchange contribution to
$\omega'$ {\em electroproduction}.  The step from photoproduction to
electroproduction is a key ingredient of the CLAS12 program, and a yield of
about $10^7$ equivalent 10 GeV photons per second on a 30 cm long liquid
hydrogen target is expected \cite{MBPC}, corresponding to about $10^5$ events
per nb per $10^7$ second year of exposure.  Thus one should at least expect
about ten thousand events of $\omega'$ electroproduction via
$\pi^0$ exchange.

The final step is extrapolation to $X(1865)$ electroproduction via $\pi^0$
exchange.  As mentioned, the price one has to pay for a process allowed for
$Q^2 > 0$ but forbidden for $Q^2 = 0$ is roughly an order of magnitude in cross
section, leading to a predicted cross section of ${\cal O}(10~{\rm pb})$,
so a signal at CLAS12 of up to a thousand events is conceivable.
A key factor signaling the electroproduction of a $0^{--}$ state will be
the vanishing of the cross section linearly as $Q^2 \to 0$. 

\section{POSSIBLE VARIANT APPROACHES} \label{sec:var}

The expected signal of ${\cal O}(10^3)$ events of $X(1865,0^{--})$
electroproduction via $\pi^0$ exchange provides some leeway when considering
possible modifications of our estimate.

\begin{itemize}
\item The ratio of  $X\to \rho\pi$ and $\omega'\to\rho\pi$ electroproduction 
events scales as the square of the ratio 
of $X$ and $\omega'$ partial widths to $\rho\pi$. The latter width seems to 
dominate a total $\omega'$ width of about 300 MeV \cite{PDG}. Thus a number of 
$X$ signal events less than ${\cal O}(10^3)$ is unlikely as long as this resonance width 
is not much smaller than 300 MeV.

\item Regge phenomenology could have been used to estimate $X(1865)$
electroproduction.  However, at small momentum transfer the exchange of an
elementary pion is almost the same as the exchange of a pion trajectory (see
Ref.\ \cite{Wang:2015kia}).

\item Other Regge trajectories, such as $a_1$ and $a_2$, could have been
considered.  However, in contrast to pion exchange, where we do see
evidence for $X$--$\rho$--$\pi$ coupling, we cannot estimate the couplings of
these other trajectories to $\rho$--$X$.  Their contribution relative to pion
exchange could be estimated by studying the energy dependence of $X$
electroproduction, which is different for pion exchange and trajectories
with intercept 1/2 such as $a_2$, and looking for evidence of the pion pole
in $t$ dependence.

\item In the absence of experimental information, we cannot estimate the
effect of a possible direct coupling of the $X$ to the proton, though if it
exists it is unlikely to interfere destructively with other sources of $X$.

\item The estimate of $a_2$ photoproduction in Ref.\ \cite{Wang:2015kia} due to
Reggeized charged pion exchange yields $\sigma(\gamma p \to a_2 n) \simeq 200$
nb at a photon energy of 10 GeV.  When extrapolating this to $e^- p \to e^-
X(1865) p$ electroproduction, note that (i) the pion-nucleon coupling for
neutral meson photoproduction via $\pi^0$ exchange is a factor of $\sqrt{2}$
less, suppressing the rate by at least a factor of two; (ii) the $|t_{\rm
min}|$ suppression factor is $\sim 0.85$; (iii) the vanishing of the cross
section at $Q^2 = 0$ imposes at least another order of magnitude suppression.
One still would expect a cross section of several nb, which is far above our
less optimistic estimate of 10 pb.  This tension will be resolved once the
$\omega'$ photoproduction and electroproduction cross sections are measured
at real or virtual photon energies around 10 GeV.

\end{itemize}

Our treatment thus may be considered as a minimal set of assumptions providing
an order-of-magnitude estimate of a signal.  We prefer to rely to the greatest
possible extent on experimental checks and to the least degree upon theoretical
calculations.  The stepwise program we have suggested provides a means to such
an estimate.

\section{SUMMARY} \label{sec:sum}

We have proposed a test for the existence of an exotic isoscalar resonance
dominating $D^0 \to \pi^+ \pi^- \pi^0$ decays.  It involves isolating
neutral-pion exchange in the electroproduction process
\beq
e^- + p \to e^- + X(1865,J^{PC}=0^{--}) + p~,
\eeq
with subsequent decay of $X(1865)$ into all three charge states of $\rho \pi$.
It is a multi-step program well suited to an intermediate-energy accelerator
such as the 12 GeV upgrade at JLAB. The steps include (i) study of $\omega(782)$
electroproduction, including isolation of the pion-exchange contribution,
(ii) a similar investigation for $\omega'$, the radial excitation of
$\omega(782)$ around 1670 MeV, and (iii) search for $X(1865)$, including the
expected vanishing of its electroproduction cross section as $Q^2 \to 0$.

\section*{ACKNOWLEDGMENTS}

We thank Mario Battaglieri, Mario Gaspero, Tim Gershon, Richard T. Jones, Brian
Meadows, Matt Shepherd, and Abi Soffer for useful discussions.  M. G. is
grateful to the CERN department of theoretical physics for its kind hospitality 
towards the completion of this work.  J.L.R.\ is grateful to the Technion for
hospitality during the inception of this work,
which was supported in part by the United States Department of Energy through
Grant No.\ DE-FG02-13ER41598, and performed in part at the Aspen Center for
Physics, which is supported by National Science Foundation grant PHY-1066293.

\section*{APPENDIX: RELATION BETWEEN $g_{\omega \pi^0 \gamma}$ AND
$g_{\omega' \pi^0 \gamma}$}

We have assumed equal differential cross sections for the pion-exchange
contribution to electroproduction of $\omega(782)$ and its presumed radial
excitation $\omega'$.  This requires the pion-photon coupling constants
for $\omega(782)$ and $\omega'$ to be the same.  We can examine the
validity of this assumption using the hadronic width of $\omega'$ and
vector meson dominance.

The decay $\omega(p) \to \pi^0(q) \gamma(k)$ may be described by the covariant
matrix element
\beq
{\cal M} = g_{\omega \pi \gamma} \epsilon_{\mu \nu \kappa \lambda}
\epsilon^\mu(p) \epsilon^\nu(k) p^\kappa k^\lambda~,
\eeq
yielding the expression
\beq
\Gamma(\omega \to \pi^0 \gamma) = \frac{g^2_{\omega \pi \gamma}p^{*3}}
{12 \pi m_\omega^2}~,~~ p^* \equiv \frac{M_\omega^2 - m_\pi^2}{2 m_\omega}~,
\eeq
where $p^* = 379.9$ MeV is the magnitude of the center-of-mass three-momentum
of either final particle.  Using values of masses, branching ratios, and
widths from \cite{PDG}, one finds $g_{\omega \pi \gamma} = 0.544$.

Taking $g_{\omega' \pi \gamma} = g_{\omega \pi \gamma}$,
and noting for $\omega' \to \pi^0 \gamma$ that $p^* = 829.5$
MeV, one finds $\Gamma(\omega' \to \pi^0 \gamma) = 1.61$ MeV.  This value is
now used to calculate $\Gamma(\omega' \to \pi^0 \rho^0)$ 
applying vector meson dominance.

The matrix element of the vector current between the vacuum and the
one-$\rho$-meson state may be parametrized as
\beq
\bra{0} V_\mu \ket{\rho^0(p)} = \epsilon_\mu(p) f_\rho m_\rho~,
\eeq
where $\epsilon_\mu$ is the $\rho$ polarization vector.  The quantity
$f_\rho$ (the $\rho$ meson decay constant) may be evaluated using the
relation
\beq
\Gamma(\rho \to e^+ e^-) = \frac{4 \pi \alpha^2 f_\rho^2}{3 m_\rho}
= (7.04 \pm 0.06)~{\rm keV}~,
\eeq
where we have neglected the electron mass and the experimental value is
that quoted in \cite{PDG}.  The result is $f_\rho = (156.4 \pm 0.7)$ MeV.
(A similar value is obtained from the decay $\tau \to \rho \nu$.)  Now
we can write
\beq
\Gamma(\omega' \to \pi^0 \gamma) = \left( \frac{e f_\rho}{m_\rho} \right)^2
\left( \frac{p^*(\omega' \to \pi^0 \gamma)}{p^*(\omega' \to \pi^0 \rho)}
\right)^3 \Gamma(\omega' \to \pi^0 \rho^0)~~,
\eeq
where $p^*(\omega' \to \pi^0 \gamma) = 829.5$ MeV, $p^*(\omega' \to \pi^0
\rho^0) = 646.3$ MeV, with the result $\Gamma(\omega' \to \pi^0 \rho^0) = 
203.8$ MeV, or, accounting also for decays to $\pi^\pm \rho^\mp$,
\beq
\Gamma(\omega' \to \pi \rho) = 611~{\rm MeV}~.
\eeq
Now, Ref.\ \cite{PDG} lists $\Gamma_{\rm tot}(\omega') = (315 \pm 35)~{\rm
MeV}$. This implies that we should take
\beq
g^2_{\omega' \pi^0 \gamma}/g^2_{\omega \pi^0 \gamma} \le (315/611) \simeq 0.5~,
\eeq
leading to a similar reduction of the $\omega'$ photoproduction cross section
as implemented in Sec.\,\ref{sec:sig}.

\end{document}